\documentclass[aps,pre,twocolumn,floatfix,amsmath,showpacs]{revtex4-1}
\usepackage{graphicx}
\usepackage{amssymb}
\usepackage{color}
\DeclareGraphicsExtensions{.pdf,.png,.jpg}
\begin{document}
\newcommand{\wq}[1]{\textcolor{blue}{#1}}

\title{Anisotropic anomalous diffusion modulated by log-periodic
oscillations
}
\author{L.~Padilla}
\email{lorenapadilla.r@gmail.com}
\author{H.~O.~M\'artin}
\email{hmartin@mdp.edu.ar}
\author{J.~L.~Iguain}
\email{iguain@mdp.edu.ar}
\affiliation{Instituto de Investigaciones F\'{\i}sicas de Mar del Plata (IFIMAR) and
Departamento de F\'{\i}sica FCEyN,\\
Universidad Nacional de Mar del Plata, De\'an Funes 3350, 7600 Mar del
Plata, Argentina}


\begin{abstract}
We introduce finite ramified self-affine substrates in two dimensions with a 
set of appropriate hopping rates between nearest-neighbor sites, 
where the diffusion of a single random walk presents an anomalous 
{\it anisotropic} behavior modulated by log-periodic oscillations. 
The anisotropy is revealed by two different random walk exponents, $\nu_x$ 
and $\nu_y$, in the {\it x} and {\it y} direction, respectively. The values of
these exponents, as well as the period of the oscillation, are analytically 
obtained and confirmed by Monte Carlo simulations.
\end{abstract}

\maketitle
\section{
Introduction
\label{intro}
}

The underlying mechanisms of anomalous diffusion on fractal structures has  
 attracted the attention of scientists for many years 
(see, for example Ref.~\cite{ale} and references therein). 
In this regard,  it has been recently found that, on some kind of self-similar 
substrates, in addition to the well-known subdiffusive behavior, the 
mean-square displacement of a random walk (RW) is modulated by logarithmic
 periodic oscillations~\cite{{woess},{lore1},{lore2}}. 
The same kind of modulation was also observed in biased diffusion on random 
systems~\cite{bernas}, earthquake dynamics~\cite{huang}, escape probabilities 
in chaotic maps~\cite{pola}, processes on random quenched and fractal 
media~\cite{kut}, diffusion-limited aggregates~\cite{sor1i}, growth 
models~\cite{huang2}, and stock markets~\cite{sor2}. There is general
agreement that this ubiquitous phenomenon appears because of an inherent 
self-similarity~\cite{dou}, responsible for a discrete scale 
invariance~\cite{sor1}. Nevertheless, this self-similarity 
has to be identified for every system.

The origin of log-periodic modulation can be 
easily determined for a minimal model of RW introduced in~\cite{lore1}.
This model, which depends on two parameters, $L \in \mathbb{N}$ 
and $0<\delta\in\mathbb{R}$, consists of a one-dimensional lattice and a 
single particle moving by jumps between nearest-neighbor (NN) sites.  
The hopping rates are defined in a way that a  region of 
size $L^n$ (with $n= 0,1,2...$) is characterized by a diffusion 
coefficient $D^{(n)}$, and the ratio between any two consecutive coefficients 
is a constant, i.~e., $D^{(n+1)}/D^{(n)}=\delta$ for all $n\in \mathbb{N}$.
As a result, the RW mean-square displacement is modulated by log-periodic 
oscillations and, both the RW exponent and the period of the oscillations
can be obtained using rather simple arguments and calculations (for more 
details, see~\cite{lore1}). 

This method can also be applied to the study of RW on a self-similar 
substrate in two dimensions. It has been shown~\cite{lore2} that,
in this case, each region of size $L^n \times L^n$ ($L$ is 
the basic length of the substrate, and $n=0, 1, 2,...$) 
is characterized by a diffusion coefficient $D^{(n)}$.
Here again a subdiffusive behavior modulated by log-periodic oscillations
 arises, because the ratio $D^{(n+1)}/D^{(n)}$ takes a constant value.
It is the  symmetry between {\it x} and 
{\it y} directions which allows the heuristic arguments used in the 
one-dimensional case to be easily generalized to calculate de values of the RW 
exponent and the period of the oscillations. 
The important point is
that, for a particle in a central square of size $L^{n} \times L^{n}$, 
the typical time to leave this square along the {\it x} direction is 
the {\it same} as that along the {\it y} direction.

In this paper we investigate single particle diffusion on self-affine 
structures. In general, the lack of symmetry between the two main directions 
($x$ and $y$) makes the analytical treatment difficult. 
However, the problem simplifies considerably for a special kind of substrate, 
that in which the space explored by a RW grows with the same anisotropy
as the substrate itself does. We study this case first.
The same kind of arguments employed to analyze
diffusion on self-similar substrates allows us to show that, in this case,
the mean-square displacement as a function of time is a power-law modulated 
by log-periodic oscillations but, in contrast with its self-similar analog,
 the specific properties of this function are now direction-dependent. 
Indeed, although the period of the modulation is isotropic, two different 
RW exponents exist, one for the displacement in the $x$ direction, 
another for the displacement in the $y$ direction. We compute analytically 
the RW exponents and the period of the modulating oscillation, and confirm 
these results by Monte Carlo simulations. 

For the sake of completeness, we then study numerically the RW behavior on a 
more general self-affine substrate. The outcomes of these simulations 
suggest that, also here, the mean-square displacements along the 
$x$ and $y$ directions, as a function of time, follow log-periodic modulated 
power-laws, which are independent of each other.

\section{Analytical Approach
\label{analy}
}
We study the behavior of a RW on two self-affine substrates, 
referred in what follows as  model I and model II. Each substrate is built in 
stages, and the result of every stage is called a {\it generation}:
a periodic array of basic or unit cells which consists of sites connected 
by bonds. We denote by $L_{x}$ and $L_{y}$ 
the linear size of the unit cell of the first generation in 
the {\it x} and  {\it y} directions, respectively. On these substrates the 
motion of a single particle occurs stochastically. At every time step, the 
particle jumps with a non-zero probability only  between NN sites 
which are connected by a bond. The details of each models are given below.

\subsection{
Model I
}

\begin{figure}[h]
\includegraphics[scale=0.5, trim=150 400 80 100]{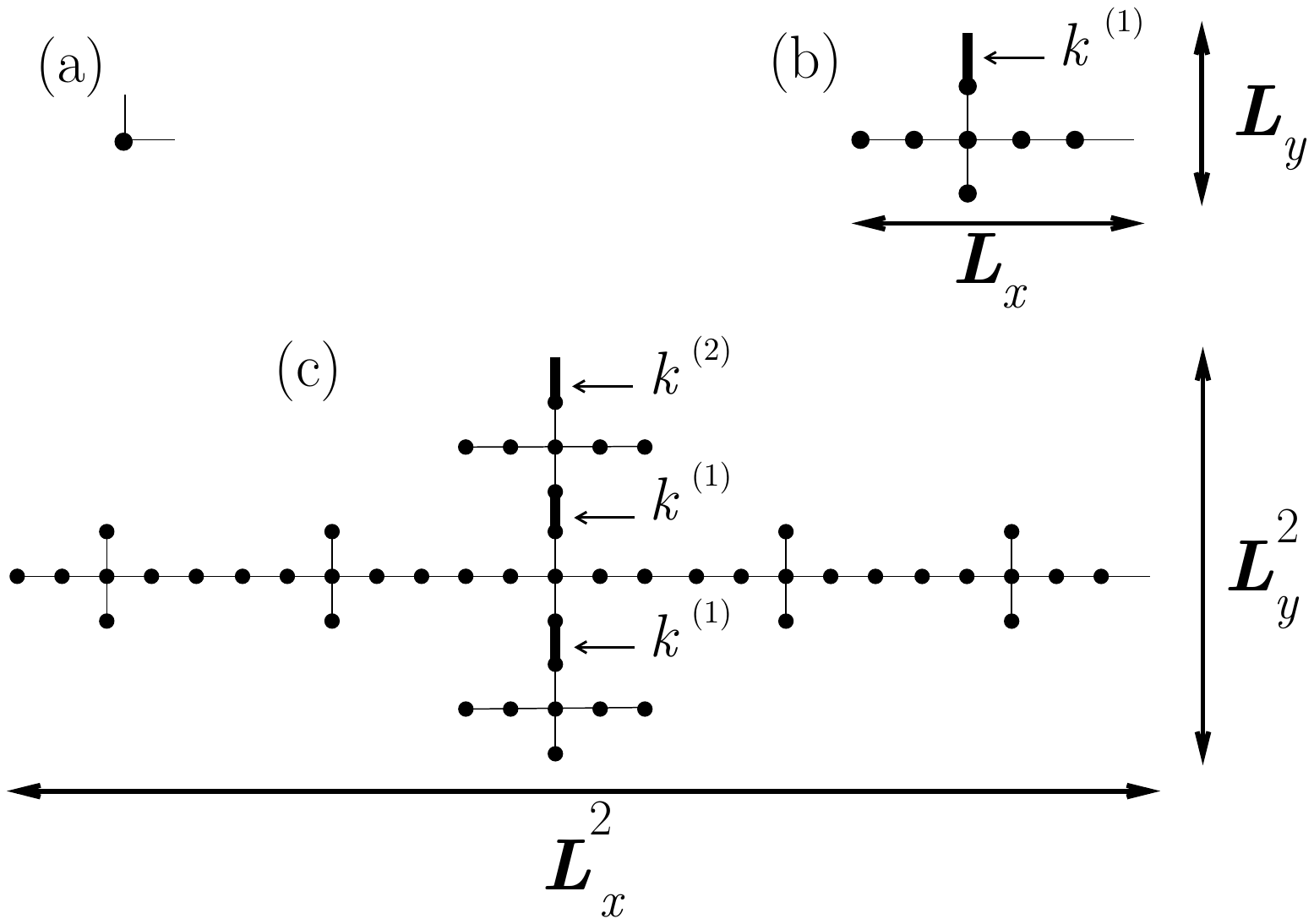}
\caption{The unit cells of model I. The zeroth, first, and second generation 
are drawn in (a), (b), and (c), respectively. The basic length-scales are
$L_x=5$ and $L_y=3$. A thin bond (thick bond) represents a hopping rate 
$k^{(0)}$ ($k^{(i)},\;\;\;i\geq 1$). More details in the text.}
\label{mo1}
\end{figure}

\begin{figure}[h]
\includegraphics[scale=0.7, trim=120 400 0 100,clip]{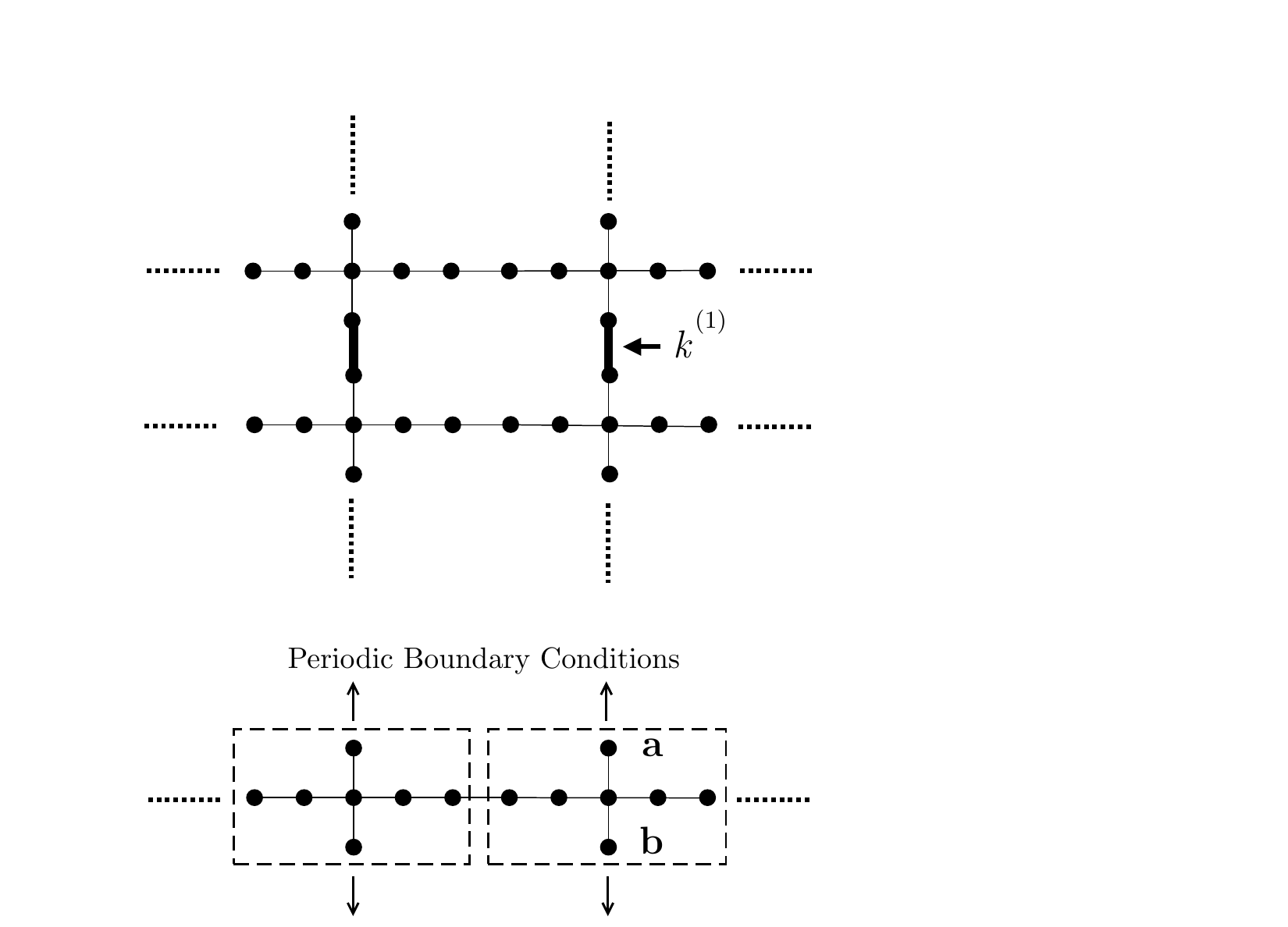}
\caption{First generation of model I. Top: the substrate built with the basic 
cell shown in Fig.\ref{mo1}(b). Bottom: the infinite one-dimensional string of 
cells used to compute the diffusion coefficient $D^{(1)}_y$. The arrows 
indicate periodic boundary condition in the {\it y} direction.
For example, if a RW at site {\bf a} ({\bf b}) jumps upward (downward), with a 
hopping rate $k^{(1)}$, it arrives at site {\bf b} ({\bf a}).
}\label{perio}
\end{figure}

The building process is illustrated in Fig.~\ref{mo1}, which shows the unit 
cell for the zeroth, first, and second generation. 
It is easy to see that, for this model, $L_{x}=5$ and $L_{y}=3$, where 
the length unit is the distance between NN sites. 
It is also apparent from this figure that the second-generation unit cell has 
linear sizes $L^{2}_{x}$ and $L^{2}_{y}$, in {\it x} and {\it y} direction, 
respectively, and is built from the first-generation one in a self-affine way. 
In general, the linear sizes of the {\it n}th-generation unit cell are 
$L^n_x$ and $L^n_y$, and the corresponding two-dimensional periodic substrate 
is obtained by connecting these cells (the first-generation substrate 
is sketched at the top of Fig.~\ref{perio})

The full self-affine substrate, we are interested in, is the result of
an infinite number of iterations.
Note that this substrate is finitely ramified, and that a region of size 
$L^n_x \times L^n_y$ can be separated from the rest by cutting four bonds.

The hopping rate between any NN connected sites in the {\it x} direction 
is always $k^{(0)}$. On the other hand, the hopping rate in the {\it y} 
direction depends on the site and on the generation. Their values are
determined by asking that the mean time to leave a $n$th-generation unit 
cell along the $x$ and $y$ directions coincide. 
We call $t^{(n)}$ this escape time.
Because of this constraint, there will be 
$n+1$ different hopping rates ($k^{(i)},\;\;\;i=0,...,n$) related to the 
$n$th generation. As an example, in Fig.~\ref{mo1} we show an schematics of the
the zeroth, first and second generation, with one, two and three kinds of hopping rates,
 respectively. In this sketch, a thin bond represents  
$k^{(0)}$, while the other hopping rates are represented by thicker bonds.
We can observe that $k^{(1)}$, appears at the top of the first generation
unit cell, and $k^{(2)}$ appears at the top of the second generation one.

We proceed now to analyze the behavior of the diffusing particle on a
$n$th-generation substrate. 
It is useful to remember that, on any periodic substrate, normal diffusion 
should be observed if time is long enough for the RW to be influenced by the 
structure periodicity. As we work with an asymmetric substrate (i.e 
$L^{n}_{x}\neq L^{n}_{y}$, for the {\it n}th-generation) we have to
consider $x$ direction and $y$ direction separately. 
For the {\it n}th-generation substrate, a diffusion 
coefficient $D^{(n)}_{x}$ ($D^{(n)}_y$) in {\it x} ($y$) direction 
can be defined through the time dependence of the mean-square 
displacement \mbox{$\Delta^2 x(t) = \langle[x(t)-x(0)]^2\rangle$}
({$\Delta^2 y(t) = \langle[y(t)-y(0)]^2\rangle$}), i.e., via the relations

\begin{equation}
\Delta^2 x(t) = 2 D^{(n)}_{x} t,
\label{evox}
\end{equation}
and
\begin{equation}
\Delta^2 y(t) = 2 D^{(n)}_{y} t,
\label{evoy}
\end{equation}
valid for a time {\it t} longer than $t^{(n)}$.

The diffusion problem is trivial on the zeroth-generation substrate. This is
a simple square lattice, and 

\begin{equation}
    D^{(0)}_{x}=D^{(0)}_{y}=k^{(0)}.
\end{equation}

The first-generation substrate (top of fig.~\ref{perio}) presents a more 
difficult task. 
However, regarding {\it x}-direction diffusion, the whole substrate and the 
string of cells displayed at the bottom of the same figure, with periodic 
boundary conditions in the {\it y} direction, lead to equivalent problems. 
We exploit this equivalence and calculate the diffusion coefficient of that 
one-dimensional array, following the steady-state method \cite{celso}.
We get

\begin{equation}
D^{(n)}_{x}=(\frac{5}{7})^{n}k^{(0)}, \;\;\;\mbox{for}\; n=0,1,2,...\;,
\label{xcoef}
\end{equation}
and thus,

\begin{equation}
 D^{(n)}_{x}/D^{(n+1)}_{x}=\delta_{x}=7/5,\;\;\;\mbox{for}\; n=0, 1, 2,...\;.
 \label{deltaDx}
\end{equation}

To find the diffusion coefficients in the case of the $y$ direction,
we divide Eq.~(\ref{evoy}) by Eq.~(\ref{evox}), imposing the 
{\it same escape time} constraint, i.e., 
$\Delta^2x(t^{(n)})=L_x^{2n}$ and $\Delta^2y(t^{(n)})=L_y^{2n}$. This leads to
 
\begin{equation}
\frac{D_y^{(n)}}{D_x^{(n)}}=\frac{L_y^{2n}}{L_x^{2n}},
\label{constraint}
\end{equation}
where the diffusion coefficients 
\begin{equation}
D^{(n)}_{y}=(\frac{9}{35})^{n}k^{(0)}, \;\;\;\mbox{for}\; n=0, 1, 2,...\;.
\label{Dy}
\end{equation}
can be obtained from (using Eq.~(\ref{xcoef}) and the 
values of $L_x$ and $L_y$). Hence, the ratio between consecutive 
coefficients is also a constant in the {\it y} direction:

\begin{equation}
  D^{(n)}_{y}/D^{(n+1)}_{y}=\delta_{y}=35/9,\;\;\;\mbox{for}\; n=0, 1, 2,...\;.
\label{deltaDy}
\end{equation}

At this stage, the model is completely defined, and the hopping rates
are obtained recursively from (\ref{deltaDx}), by using the above 
mentioned trick of converting the diffusion two-dimensional problem in a 
one-dimensional problem:

\begin{equation}
 \frac{k^{(n)}}{k^{(0)}}=[L^{n}_{x}-(L_{y}L^{n-1}_{x}-\frac{k^{(0)}}{k^{(n-1)}})]^{-1},\;\;\;
 \mbox{for}\; n=1, 2, 3,...\;.
 \label{ki}
\end{equation}

\begin{figure}[h]
\includegraphics[scale=0.85, trim=80 520 100 0,clip]{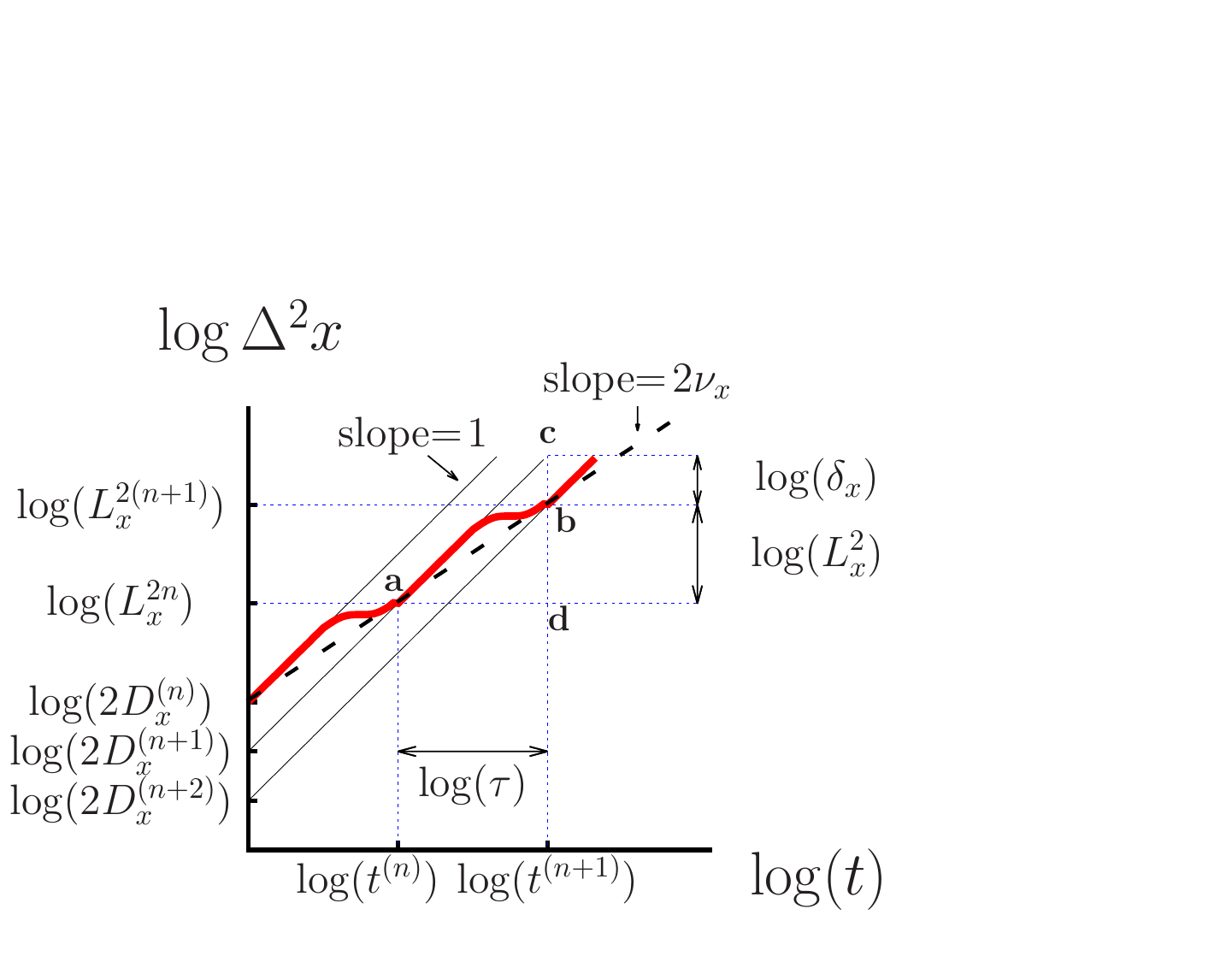}
\caption{(Color online) Schematic of the mean-square displacement in the 
{\it x} direction, as a function of the time, shown by the red (tick) curve. 
The length of the segment \textbf{bc} is 
$\log_{10}(2D^{(n)}_x)-\log_{10}(2D^{(n+1)}_x)=\log_{10}(\delta_x)$, 
because of Eq.(\ref{deltaDx}). From the slopes ($=1$) of the full straight 
lines (representing the normal diffusion behaviors, $\Delta^2x=2D^{(n)}_xt$), 
one gets that the segments \textbf{ad} and \textbf{cd} have the same 
length or, equivalently, that 
$\log_{10}(\tau)=\log_{10}(L^2_x)+\log_{10}(\delta_x)$. 
The dashed straight line represents the global power law 
$\Delta^2x\sim t^{2\nu_x}$ with  $2\nu_x=\log_{10}L^2_x/\log_{10}\tau$. Thus, 
$\nu_x=(2+\log_{10}\delta_x/\log_{10}L_x)^{-1}$. The mean-square displacement 
in the {\it y} direction exhibits an analogous behavior.}
\label{qualitative}
\end{figure}

Let us now consider a RW on the full self-affine structure. 
For a  time $t$ in the interval $[t^{(n)},t^{(n+1)}]$, the
following relations hold

\begin{equation}
L^{n}_{x}\lesssim \sqrt{\Delta^2 x(t)} \lesssim L^{n+1}_x,
\label{deltax}
\end{equation}
\begin{equation}
L^{n}_{y}\lesssim \sqrt{\Delta^2 y(t)} \lesssim L^{n+1}_y,
\label{deltay}
\end{equation}
and it will be impossible for the RW to distinguish the full 
self-affine structure from the {\it n}th-generation one. 
Thus, Eqs.~(\ref{evox}) and (\ref{evoy}) account for the RW behavior  
in that time window, and the mean-square displacement should behave 
qualitatively as sketched in Fig.~\ref{qualitative}. 
This behavior is reminiscent of single particle diffusion on a self-similar
substrate, whose mean-square displacement as a function of time obeys a 
log-periodic modulated 
power-law~\cite{lore2}. Because of the lack of symmetry between the $x$ 
and the $y$ directions, to describe diffusive behavior in the case of a 
self-affine substrate, we need not one but two functions, which we expect to 
be

\begin{equation}
\Delta^2 x(t)=C_{x}t^{2\nu_{x}}f_{x}(t)
\label{fx}
\end{equation}
and
\begin{equation}
\Delta^2 y(t)=C_{y}t^{2\nu_{y}}f_{y}(t),
\label{fy}
\end{equation}
where $C_{x}$ and $C_{y}$ are constants, $\nu_{x}$ and $\nu_{y}$ are  the RW 
exponents, and  $f_{x}(t)$ ($f_{y}(t)$) is a log-periodic function 
with period $\tau_{x}$ ($\tau_{y}$). 

The values of these quantities can be computed from the parameters
of the model, after simple geometrical analysis of Fig.~\ref{qualitative}
(see figure caption and Refs.~\cite{lore2,lore1} for further details).
The results are
 
\begin{equation}
\nu_{x} = \frac{1}{2+\frac{\log_{10} \delta_{x}}{\log_{10} L_{x}}}\;\;\; , 
\label{exponentx}
\end{equation}

\begin{equation}
\nu_{y} = \frac{1}{2+\frac{\log_{10} \delta_{y}}{\log_{10} L_{y}}}\;\;\;,
\label{exponenty}
\end{equation}

\begin{equation}
\tau_x=\delta_x L_x^2
\label{taux}
\end{equation}
and
\begin{equation}
\tau_y=\delta_y L_y^2.
\label{tauy}
\end{equation}

Note that, even when $\nu_{x}\neq\nu_{y}$, 
the period of the  modulations coincide, because of the constraint 
(\ref{constraint}), i.e., 
\begin{equation}
\tau_x=\delta_x L_x^2=\frac{D_x^{(n)}}{D_x^{(n+1)}} L_x^2 =
\frac{D_y^{(n)}}{D_y^{(n+1)}} L_y^2 = \delta_y L_y^2=\tau_y,
\label{tau}
\end{equation}
where we have also used (\ref{deltaDx}) and (\ref{deltaDy}). We call $\tau$
this period. From the equations above, the values of the period and the 
exponents are $\tau=35$, $\nu_x=0.4527$ and $\nu_y=0.3090$. 

Let us note that the average time to escape from a unit cell of the 
{\it n}th-generation is $t^{(n)}=\tau^n$,
which means that the relations (\ref{deltax}) and (\ref{deltay}) hold for,

\begin{equation}
\tau^n\lesssim t \lesssim \tau^{n+1}.
\label{tao}
\end{equation}
Then, when the RW leaves the initial region, of size 
$L^{n}_x \times L^{n}_y$, to enter the next one, of size 
$L^{n+1}_x \times L^{n+1}_y$, the length-width ratio 
($L^n_x/L^n_y$) is increased by an anisotropic factor $a=L_x/L_y$, while 
the average time increases from 
$t$ to $\tau t$. On the other hand, according to (\ref{fx}) and 
(\ref{exponentx}), the corresponding mean square displacements   
are related by $\Delta^2x(\tau t)=L^2_x\Delta^2x(t)$ and
$\Delta^2y(\tau t)=L^2_y\Delta^2y(t)$. Therefore, in this transition, 
the ratio $\sqrt{\Delta^2x/\Delta^2y}$ is also increased by a factor 
$a$; i.e., the space explored by the RW grows with the same 
anisotropy as the substrate where the diffusion takes place.

\subsection{
Model II
}

\begin{figure}[h]
\includegraphics[scale=0.7, trim=120 450 0 120,clip]{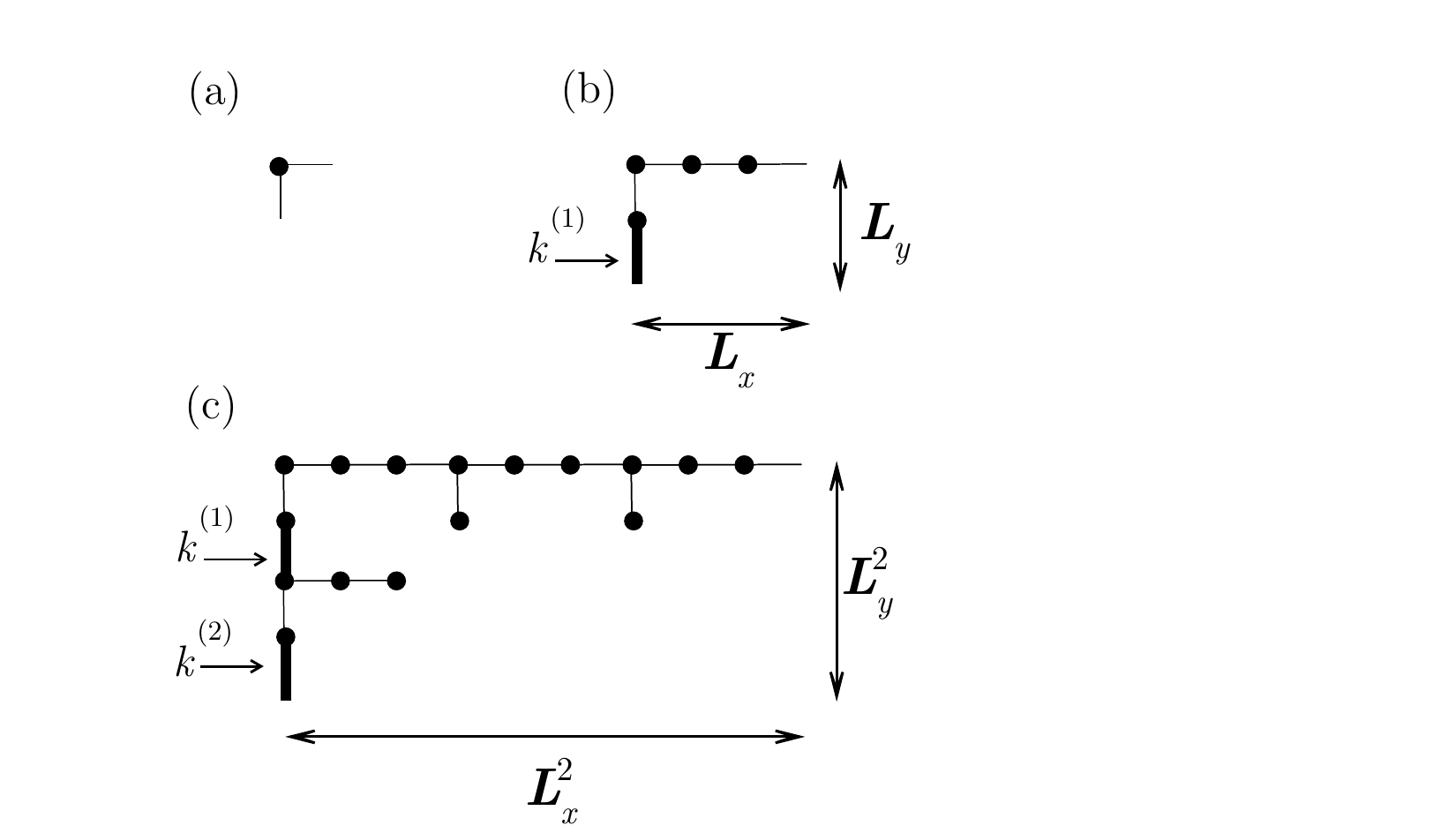}
\caption{The unit cells of model II. The zeroth, first, and second 
generations are shown in (a), (b), and (c), respectively. $L_{x}=3$ and 
$L_{y}=2$.}
\label{mo2}
\end{figure}

For this model, the unit cells for the zeroth, first and second generation are 
shown in Fig.~\ref{mo2}. The full self-affine substrate is here also obtained 
when the generation order goes to infinity. The linear sizes of the 
{\it n}th-generation unit cell are $L^{n}_{x}$ and $L^{n}_{y}$, with 
$L_{x}=3$ and $L_{y}=2$.

The diffusion of a single particle is analyzed as on model I. That is, we  
reformulate the two-dimensional RW problem on a one-dimensional array 
and compute the diffusion coefficients following the steady-state 
method~\cite{celso}. 

For the $n$th generation, we obtain

\begin{equation}
D^{(n)}_{x}=(\frac{3}{4})^{n}k^{(0)},\;\;\;\mbox{for}\; n=0, 1, 2,...\;,
\end{equation}
and thus,

\begin{equation}
D^{(n)}_x/D^{(n+1)}_x=\delta_x=4/3,\;\;\;\mbox{for}\; n=0, 1, 2,...\;.
\label{cond2}
\end{equation}
In average, the time to leave a $n$th-generation unit cell 
along the $x$ direction becomes the same as that along the $y$ direction if

\begin{equation}
D^{(n)}_{y}=\frac{k^{(0)}}{3^n},\;\;\;\mbox{for}\; n=0, 1, 2,...\;,
\label{Dy_2}
\end{equation}
which implies

\begin{equation}
D^{(n)}_y/D^{(n+1)}_y=\delta_y=3,\;\;\;\mbox{for}\; n=0, 1, 2,...\;.
\label{deltaDy_2}
\end{equation}

The $k^{(i)}$'s, coming from (\ref{Dy_2}), are again computed from 
(\ref{ki}) (with $L_x=3$ and $L_y=2$). Furthermore, 
in spite of the differences between model I and model II, the qualitative 
behavior sketched in Fig.~\ref{qualitative} we expect to be valid 
for both models. Therefore, the  RW exponents $\nu_x$ and $\nu_y$, and the 
period $\tau$ are given by (\ref{exponentx}), (\ref{exponenty}) 
and (\ref{tau}); with the values  $\nu_x=0.4421$, $\nu_y=0.2789$ and $\tau=12$.

\section{
Numerical Results
\label{nume}
}

To test the predictions outlined above, we perform standard RW Monte Carlo 
simulations,  on a {\it n}th-generation unit cell for each model. 
In model I (II) every RW starts at the center of symmetry of the cell 
(at the top-left most site). The value of {\it n} is 
always chosen large enough to prevent the RWs from reaching the cell borders 
(the bottom and right cell borders) during the simulation. Working on this 
cell is thus equivalent to working with the full self-affine structure. In all 
simulations 
the hopping rate
$k^{(0)}$ is set to $1/4$, and the other $k^{(i)}$'s ($i\geq1$) are 
obtained from (\ref{ki}). After every Monte Carlo step, the time
is increased by $\Delta t=1$ 

\begin{figure}[h]
\includegraphics[width=\linewidth, trim=0 80 0 0,clip]{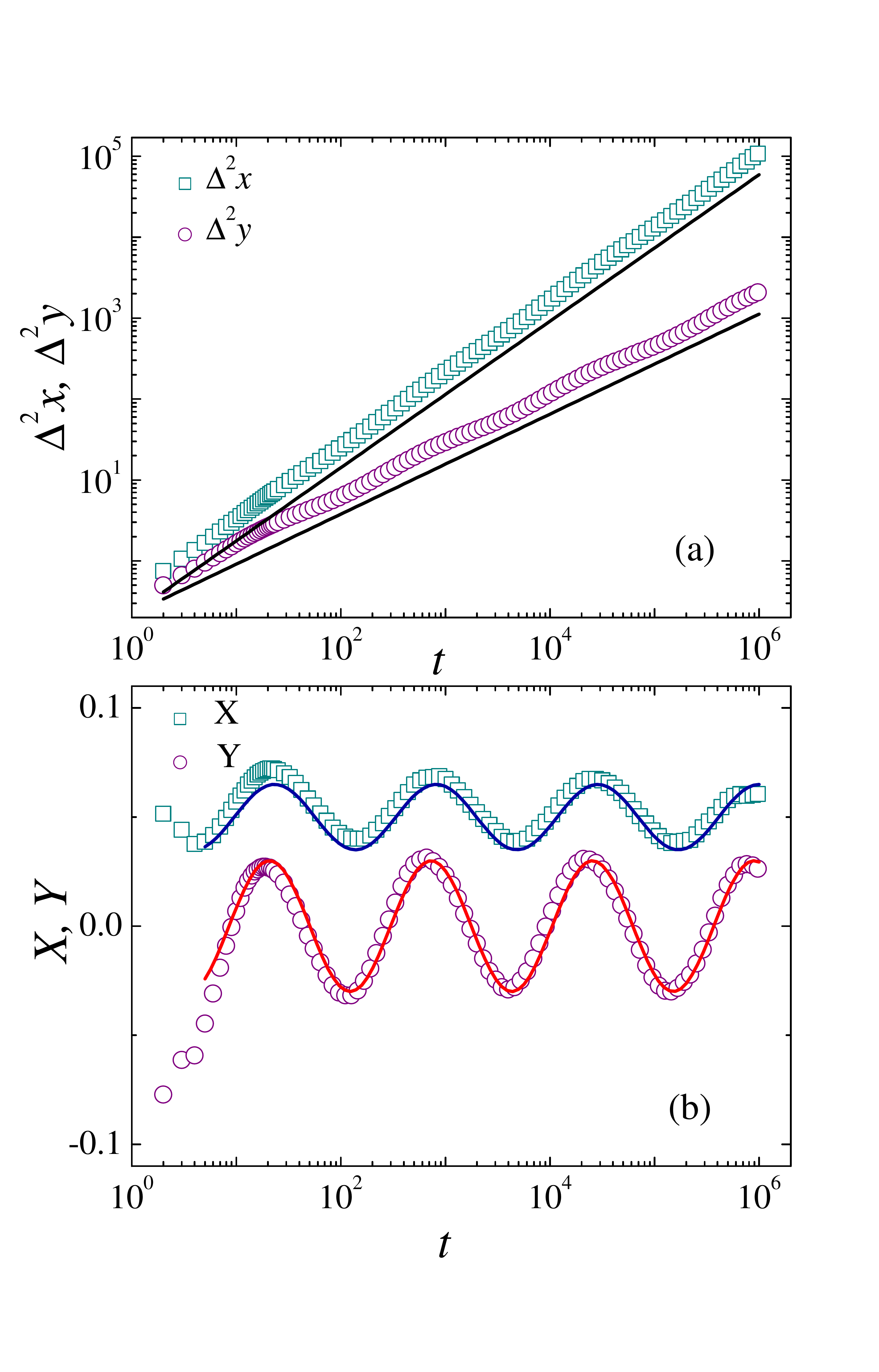}
\caption{(Color online) (a): The mean-square displacements $\Delta^2 x $ 
(green squares), and $\Delta^2 y $ (purple circles)
 as functions of time for Model I. The upper straight line
 has a slope $2\nu_x$, with $\nu_x=0.4527$  obtained from 
Eq.~(\ref{exponentx}). The lower straight line has a slope $2\nu_y$, 
with $\nu_y=0.3090$  obtained from Eq.~(\ref{exponenty}).
(b): $X=\log_{10}(\Delta^2x/A_xt^{2\nu_x})$ vs.
 $\log_{10} t$  ($Y=\log_{10}(\Delta^2y/A_yt^{2\nu_y})$ vs $\log_{10}(t)$) for 
the same data. $A_x$ and $A_y$ are a properly chosen constants. The curves
represent the first-harmonic approximations 
$B_x\sin [2\pi \log_{10}(t)/\log_{10}(\tau)]+\alpha$ (blue-upper) and
  $B_y\sin [2\pi \log_{10}(t)/\log_{10}(\tau)]+\beta$ (red-lower).
The period $\tau$ is given by Eq.(\ref{tau}). $B_x$, $B_y$, $\alpha$ 
and $\beta$ are fitted constants.}
\label{modI_lineal}
\end{figure}

With the numerical results of model I, in Fig.~\ref{modI_lineal}-(a) we have 
plotted the mean-square displacement along the main directions.
We see in these plots that both $\Delta^2x(t)$ and $\Delta^2y(t)$ 
are well described by modulated power laws. The upper and lower straight lines 
have slopes $2\nu_x$ and $2\nu_y$, respectively. They are drawn to guide 
the  eyes, using the analytical values of the RW exponents. The log-periodicity of
the modulations can be better observed in Fig.~\ref{modI_lineal}-(b), where 
we have plotted 
$\log_{10}(\Delta^2x/A_xt^{2\nu_x})$ and $\log_{10}(\Delta^2y/A_yt^{2\nu_y})$ 
against $\log_{10}(t)$, using the same data as in the part (a). 
$A_x$ ($A_y$)  is a constant chosen to have the oscillations in the $x(y)$ 
direction centered around $0.00$ ($0.05$). The continuous lines are of the form 
$B\sin (2\pi \log_{10}(t)/\log_{10}(\tau)+\alpha)$,  i.e., the first-harmonic 
approximation of a periodic function with period $\log_{10}(\tau)$, where
 $B$ and $\alpha$ are fitted  parameters and $\tau=35$ (the above given analytical period). It is clear from this figure that the theoretical 
predictions (Eq.(\ref{exponentx}), (\ref{exponenty}) and (\ref{tau})) are 
consistent with the numerical findings.

\begin{figure}[h]
\includegraphics[width=\linewidth]{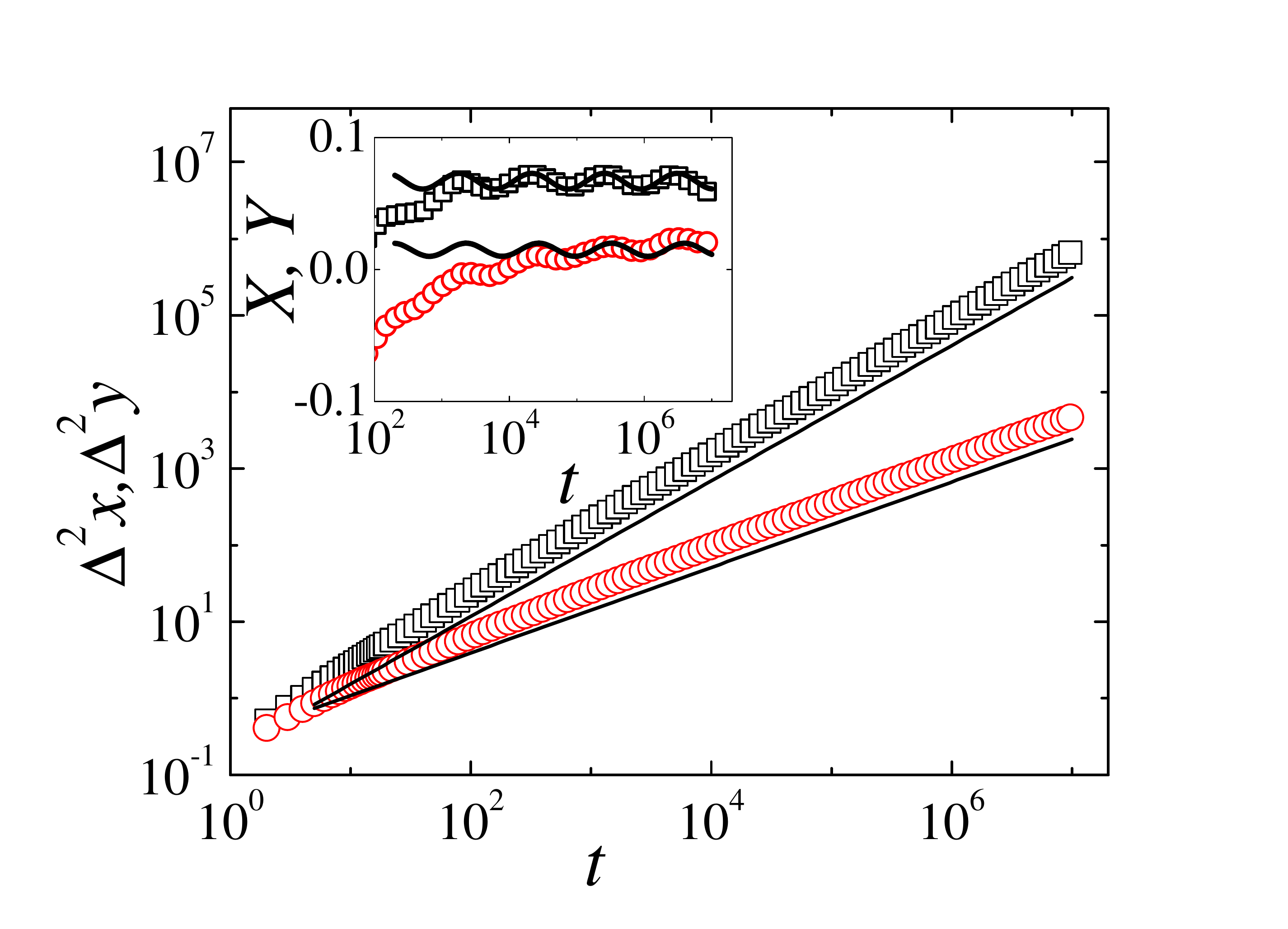}
\caption{(Color online) The mean-square displacement $\Delta^2x$ (black 
squares) [$\Delta^2y$ (red circles)] versus time for model II. The top straight
 line has a slope $2\nu_x=0.8842$, and the lower straight line has a slope 
$2\nu_y=0.5579$. Both exponents are obtained from Eq.(\ref{exponentx}) 
and (\ref{exponenty}). The inset are plots of 
$X=\log_{10}\Delta^2x/A_xt^{2\nu_x}$ (black squares) and 
$Y=\log_{10}\Delta^2y/A_xt^{2\nu_y}$ (red circles) against $t$, for the same 
data. The curves were obtained as in Fig.~\ref{modI_lineal}, with the period 
$\tau$ calculated from Eq.(\ref{tau}).
}
\label{modII}
\end{figure}

The corresponding numerical results for model II are shown in
 Fig.~\ref{modII}. Note that, also for this model, at long times, the 
mean-square displacement as a function of time is well described by 
modulated power laws. To better appreciate the log-periodicity of the 
modulation, we have plotted $\log_{10}(\Delta^2x/A_xt^{2\nu_x})$  
vs.~$\log_{10}(t)$ and 
$\log_{10}(\Delta^2y/A_yt^{2\nu_y})$  vs $\log_{10}(t)$ in the inset of this 
figure.
The fitting curves are of the form 
$D\sin (2\pi \log_{10}(t)/\log_{10}(\tau)+\alpha)$, with the analytical value 
$\tau=12$. The agreement between 
analytical and numerical results is also good.
\begin{figure}[h]
\includegraphics[width=\linewidth]{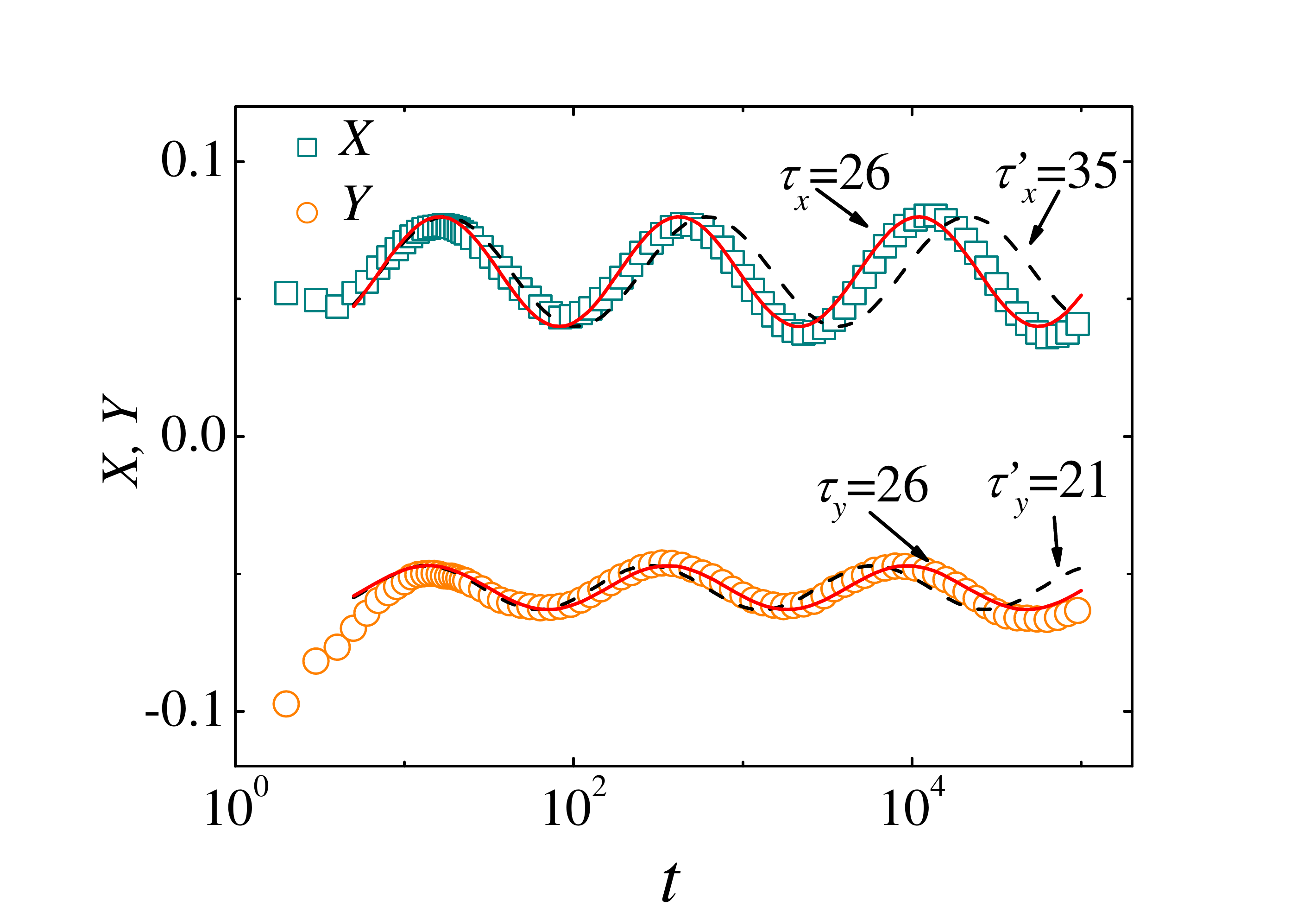}
\caption{(Color online) Scaled mean-square displacements for Model III.  
Plot of 
$X=\log_{10}\Delta^2 x / C_xt^{2\nu_x}$ vs.
 $\log_{10} t$ (green squares) and $Y=\log_{10}\Delta^2 y / C_yt^{2\nu_y}$ vs.
 $\log_{10} t$ (orange circles), using numerical data ($C_x$ and $C_y$ are  
properly chosen constants, see the text). The
full lines represent the first-harmonic approximations
$A_x \sin(2\pi \log_{10} t)/\log_{10}(\tau)+\alpha)$ (upper) and 
$A_y \sin(2\pi \log_{10} t)/\log_{10}(\tau)+\beta)$ (lower) of $X$ and $Y$, 
respectively.
Here, $\tau=26$, $A_x$, $A_y$, $\alpha$ and $\beta$ are fitted constants. 
The uppers dashed line represents the first-harmonic approximation
of $X$, with period $\tau'_x=35$. 
The lower dashed line represents the first-harmonic approximation
of $Y$, with period $\tau'_y=21$
}
\label{sinbarreraI}
\end{figure}

We consider now a substrate (model III) which consists of the full self-affine 
structure of model I but with the same 
hopping rate $k^{(0)}$ between any pair of connected NN sites. 
For this model, the average time to leave a $n$-generation unit
cell along the $x$ direction is different from that along the $y$ direction.
It may occur that 
$L^{n}_{x}\lesssim \sqrt{\Delta^2 x(t)} \lesssim L^{n+1}_x$, 
$L^{m}_{y}\lesssim \sqrt{\Delta^2 y(t)} \lesssim L^{m+1}_y$, for a given time
$t$ and $m\neq n$, which, in other words means that, near $t$, the RW behaves 
as in the $n$-generation substrate, regarding the $x$ direction, but as in 
the $m$-generation substrate, regarding the $y$ direction.
Thus, we cannot expect the heuristic arguments in the previous section continue
to be valid and we have then to study the problem numerically.

The logarithm of the scaled mean-squared 
displacements (in the {\it x} and {\it y} directions), i.e., 
$\Delta^2x/(A_xt^{2\nu_x})$ and $\Delta^2y/(A_yt^{2\nu_y})$  are plotted in 
Fig.~\ref{sinbarreraI} as a function of the logarithm of time. The RW 
exponents $\nu_x=0.4373$ and $\nu_y=0.3859$ in this figure are fitted 
values . Let us 
note that  $\nu_x$ is different from $\nu_y$, and that the data of 
Fig.~\ref{sinbarreraI} strongly suggest that the modulation have the same 
period $\tau$ in both directions.
As expected, the numerical values of these parameters are not in agreement 
with Eqs.(\ref{exponentx}), (\ref{exponenty}), and (\ref{tau}).
We would like to remark that if we used Eqs.(\ref{exponentx}) and 
(\ref{exponenty}) 
(with $\delta_x=7/5$ and $\delta_y=7/3$ resulting from the new hopping rates)
we would get
 the RW exponents $\nu'_x=0.4527$ and $\nu'_y=0.3609$, which,  in turn would lead
to the periods $\tau'_x=L^{1/\nu'_x}_x=35$ and $\tau'_y=L^{1/\nu'_y}_y=21$;
different of each other (see caption of Fig.~\ref{qualitative} for the equation
$\tau=L^{1/\nu}$). Note that the numerical value of $\nu_x=0.4373$ 
($\nu_y=0.3859$) is smaller than $\nu'_x$ (larger than $\nu'_y$), 
and the numerical value period $\tau$ is in the range $[\tau'_y, \tau'_x]$
($\tau\cong26$).
For model III, the Eqs. (\ref{deltax}), (\ref{deltay}) and (\ref{tao}) do not hold because, in average, the RW reaches the top or bottom border of
the $n$-generation unit cell before reaching the right or left border of
the same cell. In the case of model I, this is avoided by properly modifying
some hopping rates in every generation. The diffusion spread in the 
$y$ direction is thus slowed
down ($k^{(n)}<k^{(n-1)}$, see Eq.(\ref{ki})), and the 
horizontal and vertical cell borders are, in average, simultaneously reached.

\section{
Conclusions and discussion
\label{conclu}
}
We have studied  the problem of single particle diffusion
on a finitely ramified self-affine structure in two dimensions. 
For a special kind
of models, for which the ratio between the $x$ and $y$ mean-square
displacements matches the structure anisotropy, we argue that  
the RW exponent in the {\it x} direction $\nu_x$  is different from that in the
{\it y} direction $\nu_y$, and that the global subdiffusive behavior is
modulated by log-periodic oscillations with a period $\tau$ which does
not depend on the direction. The arguments employed in this work
allow the main properties of the particle mean-square displacement
to be obtained as a function of model parameters.
Because our arguments are somehow heuristic, MC 
simulations using two models, I and II, were also carried out.
The numerical results confirm our theoretical predictions.

For the rest of the self-similar systems, our conclusions are more 
limited, due to the lack of suitable analytical methods
and that the RW explores the space with an anisotropy different from 
that of the substrate.
The results of the MC simulations performed using one of these models (III),
show (within the accuracy of the simulation) that, also in this case
 $\nu_x\neq\nu_y$ and  the RW mean-square displacement
is modulated by log-periodic oscillations with an isotropic period.
However, we cannot guarantee that this behavior will hold in the limit 
of an arbitrary long time; that is why we have introduced models I and II.
Let us finally note that the extension of our analytical results
to other values of $L_x$ and $L_y$ is straightforward.

\begin{center}
\textbf{ACKNOWLEDGMENTS}
\end{center}

This work was supported by the Universidad Nacional de Mar del Plata and the Consejo Nacional de Investigaciones Cient\'{\i}ficas y T\'ecnicas-CONICET-(PIP 0041/2010-2012).

\end{document}